\documentstyle[12pt,titlepage,epsbox]{article}
\begin{document}
\pagestyle{empty}
\newcommand{\hf}{\hfill}
\newcommand{\lapproxeq}{\lower
.7ex\hbox{$\;\stackrel{\textstyle <}{\sim}\;$}}
\baselineskip=0.212in

\begin{flushleft}
\large
{SAGA-HE-74-94
   \hfill December 19, 1994}  \\
\end{flushleft}

\vspace{1.5cm}

\begin{center}

\LARGE{{\bf Numerical Solution}} \\

\vspace{0.3cm}
\LARGE{{\bf of Altarelli-Parisi Equations}} \\

\vspace{1.5cm}

\Large
{R. Kobayashi$^*$, M. Konuma, S. Kumano, and M. Miyama $^\star$ }   \\

\vspace{1.0cm}

\Large
{Department of Physics}         \\

\vspace{0.1cm}

\Large
{Saga University}      \\

\vspace{0.1cm}

\Large
{Saga 840, Japan} \\

\vspace{1.5cm}

\large
{Talk given at the workshop on} \\

\vspace{0.3cm}

{``From Hadronic Matter to Quark Matter:} \\

\vspace{0.3cm}

{ Evolving View of Hadronic Matter"} \\

\vspace{0.7cm}

{YITP, Kyoto, Japan,  October 30 -- November 1, 1994}  \\

\end{center}

\vspace{1.1cm}
\vfill

{\rule{6.cm}{0.1mm}} \\

\vspace{-0.4cm}

\normalsize
{* Speaker}  \\

\vspace{-0.2cm}
\normalsize
{$\star$ Email: kobar, konumam, kumanos,
                or 94sm10@cc.saga-u.ac.jp}  \\

\vspace{-0.2cm}
\noindent
\normalsize
{Our group activities are listed
 at ftp://ftp.cc.saga-u.ac.jp/pub/paper/riko/quantum1} \\
\vspace{-0.6cm}

\noindent
{or at
http://www.cc.saga-u.ac.jp/saga-u/riko/physics/quantum1/structure.html.} \\

\vspace{-0.2cm}
\hfill
{to be published in Prog. Theor. Phys. Supplement}

\vfill\eject
\pagestyle{plain}
\begin{center}

\Large
{Numerical Solution of Altarelli-Parisi Equations} \\

\vspace{0.5cm}

{R. Kobayashi $^*$, M. Konuma, S. Kumano, and M. Miyama $^\star$ } \\

{Department of Physics}    \\

{Saga University}      \\

{Saga 840, Japan} \\

\vspace{0.7cm}

\normalsize
Abstract
\end{center}
\vspace{-0.46cm}

\vspace{+0.3cm}

We discuss numerical solution of Altarelli-Parisi equations
in a Laguerre-polynomial method and in a brute-force method.
In the Laguerre method, we get good accuracy by taking
about twenty Laguerre polynomials in the flavor-nonsinglet case.
Excellent evolution results are obtained in the singlet case
by taking only ten Laguerre polynomials.
The accuracy becomes slightly worse in the small and large $x$ regions,
especially in the nonsinglet case. These problems could be implemented
by using the brute-force method; however, running CPU time
could be significantly
longer than the one in the Laguerre method.

\vspace{0.8cm}

\noindent
{\Large\bf 1. Introduction}

\vspace{0.2cm}

In order to investigate nucleon substructure,
we scatter high-energy electrons or muons from
the proton target.
Its cross section is expressed
in term of two unpolarized structure functions $F_1$ and $F_2$,
which depend on two kinematical variable $x$ and $Q^2$.
If the nucleon consists of free partons,
the structure functions are independent of the variable $Q^2$.
This assumption is called Bjorken scaling hypothesis.
However, it is well known that the structure functions
do depend on $Q^2$ experimentally and theoretically.
This fact is called ``scaling violation", which is
important in testing perturbative QCD.
The scaling violation results from
the fact that high-energy photons probe minute
$q\bar q$ clouds around a quark.

An intuitive way to describe the phenomena
is to use the Altarelli-Parisi equation.
The flavor-nonsinglet Altarelli-Parisi equation
is given by
$$
{d\over d\ln Q^2} ~q_{_{NS}}(x,Q^2)
={\alpha_s(Q^2)\over 2\pi}\int_x^1{dy\over y} ~
P_{_{NS}}\biggl({x\over y}\biggr) ~
q_{_{NS}}(y,Q^2) ~~~,
\eqno{(1.1)}
$$
where $q_{_{NS}}(x,Q^2)$ is a nonsinglet quark distribution,
$\alpha_s(Q^2)$ is the running coupling constant,
and $P_{NS}(x)$ is a nonsinglet splitting function.
The above equation describes the process that
a quark with the nucleon's momentum fraction $y$ radiates
a gluon and it becomes a quark with the momentum fraction $x$.
The splitting function $P_{NS}(z)$ determines
the probability for a quark
radiating a gluon such that the quark momentum is reduced
by the fraction $z$.
Next-to-leading-order corrections to
the Altarelli-Parisi equation are included
in the coupling constant $\alpha_s$
and in the splitting function $P_{NS}$.
The leading order and the next-to-leading order are abbreviated to
LO and NLO respectively throughout this paper.

The singlet Altarelli-Parisi equations are coupled
integrodifferential equations:
$$
\hspace{-0.9cm}
{d\over d\ln Q^2} q_{_{S}}(x,Q^2)={\alpha_s\over 2\pi}
\int\limits_x^1{dy \over y}   ~\left[~
P_{qq}\left( {x\over y}\right) q_{_{S}}(y,Q^2)   ~+~
P_{qG}\Biggl({x\over y}\Biggr) G(y,Q^2)          ~\right] ~~~,
$$
$$ {d\over d\ln Q^2} G (x,Q^2)={\alpha_s\over 2\pi}\int\limits_x^1
{dy \over y}                  ~\left[~
P_{Gq}\Biggl( {x\over y}\Biggr) q_{_{S}}(y,Q^2)  ~+~
P_{GG}\Biggl({x\over y}\Biggr) G(y,Q^2)          ~\right]   ~~~,
\eqno{(1.2)}
$$
where $q_{_{S}}(x,Q^2)$ is the singlet quark distribution
$
\displaystyle{
q_{_{S}} \equiv \sum_{i=1}^{N_f}(q_i+\bar{q}_i)
}
$
and $G(x,Q^2)$ is the gluon distribution.
$P_{qq}(x)$, $P_{qG}(x)$, $P_{Gq}(x)$, and $P_{GG}(x)$
are splitting functions.

Because the Q$^2$ evolution equations are very important for testing
perturbative QCD and are often used theoretically and experimentally,
it is worth while having a computer program of solving it
accurately without consuming much computing time.
Our research purpose is to create a useful computer program
for numerical solution of the Altarelli-Parisi equations.
There are several methods in solving the equations; however,
a Laguerre-polynomial method is considered to be very effective
among them \cite{LAG}.
In Ref. \cite{KKK}, we provide the Laguerre method program LAG2NS,
which deals with the Q$^2$ evolution of a nonsinglet distribution
in the next-to-leading order.
In this paper, we show Laguerre-method results in Ref. \cite{KKK} and
compare them with results in a brute-force method \cite{BRUTE}.

In section 2, we explain the Laguerre method for solving the evolution
equations with NLO corrections.
The brute-force method is discussed in section 3 and
conclusions are given in section 4.

\vspace{1.0cm}
\noindent
{\Large\bf 2. Laguerre-polynomial method}
\vspace{0.4cm}

\noindent
{\bf 2.1 Nonsinglet distribution}
\vspace{0.3cm}

The Laguerre-polynomial method is considered to be effective
in convergence and in computing time.
This is because the first several polynomials resemble
parton distributions.
Numerical solution for the evolution equation including
the NLO corrections is discussed in the following.
We refer the reader to the papers
in Refs. \cite{LAG,KKK} for more complete account of
the Laguerre method.
The NLO running coupling constant is used in this paper
unless we specify LO.
The renormalization scheme is $\overline{MS}$.

We first define the variable $t$ and the evolution operator $E(x,t)$ by
$
\displaystyle{
t = -{2\over \beta_0}\ln
\biggl[{\alpha_s(Q^2)\over \alpha_s(Q_0^2)
}\biggr]
}
$
and
$$
\widetilde{q}_{_{NS}}(x,t) =
\int_x^1{dy\over y} ~E({x\over y},t) ~\widetilde{q}_{_{NS}}(y,t=0) ~~~,
\eqno{(2.1)}
$$
where $\widetilde q$ is given by $\widetilde q(x)=xq(x)$.
The function $E(x,t)$ describes the Q$^2$ evolution
from the initial distribution at Q$_0^2$ ($t=0$)
to the one at Q$^2$.
Substituting Eq. (2.1) into the Altarelli-Parisi equation
in Eq. (1.1), we find that the evolution operator satisfies
$$
{d\over dt} ~ E(x,t)=\int_x^1{dy\over y} ~
 \biggl[\widetilde{P}_{NS}^{(0)}({x\over y})
+{\alpha_s(t)\over 2\pi}R({x\over y})\biggr]~ E(y,t)  ~~~,
\eqno{(2.2)}
$$
where $\widetilde P(x)=xP(x)$ and $R(x)$ are given by
$
\displaystyle{
R(x)= \widetilde{P}_{NS}^{(1)}(x)-{\beta_1\over 2\beta_0}
\widetilde{P}_{NS}^{(0)}(x).
}
$
$P_{NS}^{(0)}(x)$ and $P_{NS}^{(1)}(x)$ are
LO and NLO splitting functions.
The function $E(x,t)$ is split into LO and NLO contributions by
$
\displaystyle{
E(x,t)=E^{(0)}(x,t)+{\alpha_s(0)\over 2\pi}E^{(1)}(x,t) .
}
$
Substituting this into Eq. (2.2) and
considering that the LO evolution operator $E^{(0)}(x,t)$
satisfies the LO Altarelli-Parisi equation,
we find that the LO and NLO evolution operators
are not independent and they are related by
$$
E^{(1)}(x,t)={2\over \beta_0}
\biggl[1-{\alpha_s(t) \over \alpha_s(0)}\biggr]
\int_x^1 {{dy} \over y}  ~R\biggl({x \over y}\biggr)~ E^{(0)}(y,t)  ~~~.
\eqno{(2.3)}
$$

The variable $x'$ is introduced by $x^\prime = -\ln x$ ($0\le x\le 1$)
in order to use the Laguerre polynomials, which are defined in the region
$0\le x'< \infty$.
Using the variable $x'$, we expand the functions
in terms of the Laguerre polynomials:
$\displaystyle{f (x=e^{-x^\prime})=\sum_n f_n L_n(x^\prime)}$.
Then, the problem is to obtain the expansion coefficients
$E_n$ in terms of the coefficients $P_n$ of the splitting function.
The LO evolution operator is given in Ref. \cite{LAG} by
$\displaystyle{E_n^{(0)}(t)=e^{P_0t}
\sum_{k=0}^{N_{max}}{t^k\over k!}B_n^k}$,
where $B_n^k$ is calculated by the recurrence relation
$\displaystyle{B_n^{k+1}=}$
$\displaystyle{\sum_{i=k}^{n-1} (P_{n-i}-P_{n-i-1}) B_i^k}$
with
$\displaystyle{B_i^0=1,\ B_i^1=\!\sum_{j=1}^i (P_j-P_{j-1}),
\ B_0^k=B_1^k= \cdots=B_{k-1}^k=0}$.
The NLO evolution operator is given by
using the calculated $E_n^{(0)}(t)$ and Eq. (2.3):
$$
E_n^{(1)}={2\over \beta_0}
\biggl[ 1- {{\alpha_s(t)} \over {\alpha_s(0)}} \biggr]
\sum_{i=0}^n(R_{n-i}-R_{n-i-1})E_i^{(0)}(t)    ~~~,
\eqno{(2.4)}
$$
where $R_i$ are the Laguerre coefficients of $R(x)$, and
$R_{-1}$ is defined as $R_{-1}=0$.
Using these results, we finally obtain
$$
\widetilde{q}_{_{NS}}(x,t)=\sum_{n=0}^{N_{max}} \sum_{m=0}^n
\biggl[E_{n-m}(t)-E_{n-m-1}(t)\biggr]
L_n(- \ln x)\widetilde{q}_{_{NS},m}(t=0) ~~~,
\eqno{(2.5)}
$$
with $\displaystyle{E_n(t)=E_n^{(0)}(t)
      +{{\alpha_s(0)} \over {2\pi}} E_n^{(1)}(t)}$.
In this way, the original integrodifferential equation is reduced
to a sum of finite number of Laguerre expansion coefficients
and the Laguerre polynomials.

We have discussed the $Q^2$ evolution
of a nonsinglet quark distribution.
$Q^2$ evolution of a nonsinglet structure function
is calculated in a similar way.
The only modification is to
take into account NLO corrections
from the coefficient function.
A nonsinglet structure function $F_{NS}(x,Q^2)$ is expressed
as a convolution of
the corresponding nonsinglet quark distribution $q_{_{NS}} (x,Q^2)$
and the coefficient function $C_{NS} (x,\alpha_s)$ as \cite{HW}
$$
F_{NS}(x,Q^2)=\int_x^1 {{dy} \over y} ~ C_{NS} ({x \over y},\alpha_s) ~
                                       q_{_{NS}} (y,Q^2)  ~~~.
\eqno{(2.6)}
$$
$C_{NS}$ is given by
$
\displaystyle{
C_{NS} (x,\alpha_s)=\delta (1-x)
                    +{{\alpha_s} \over {4\pi}} B^{NS}(x)   ,
}
$
where $B^{NS}(x)$ is the NLO correction.
We show examples of our evolution results \cite{KKK} in the following.

\vspace{0.2cm}
\noindent
\parbox{7.5cm}
{
\hspace{0.6cm}
\baselineskip=0.212in
The initial distribution is given by
the HMRS-B valence-quark distribution at $Q^2$=4 GeV$^2$ \cite{HMRS}.
This distribution is evolved to the one
at $Q^2$=200 GeV$^2$.
Evolved distributions are shown in Fig. 1.
We find that accurate results are obtained by
taking about twenty Laguerre polynomials; however,
the accuracy becomes slightly worse in the very small
$x$ region ($x<0.001$) and in the very large $x$ region ($x>0.9$).
 } \ \hspace{0.8cm} \
\parbox{6.0cm}
{
\vspace{0.4cm}
\epsfile{file=fig1.eps,width=5.0cm}
\par\vspace{0.3cm}\hspace{-0.4cm}
Fig. 1 ~$Q^2$ evolution of $x(u_v+d_v)$.
}

\vspace{0.2cm}
\noindent
\parbox{7.5cm}
{
\hspace{0.6cm}
\baselineskip=0.212in
Our evolution results are compared with the CDHSW
neutrino data \cite{NEUTRINO} in Fig. 2.
The CCFR $F_3(x)$ distribution at $Q^2$=3 GeV$^2$
\cite{CCFR} is chosen as the initial distribution
in both LO and NLO cases. We employ $\Lambda$=0.21 GeV and
four flavors for calculating the evolution.
The NLO contributions are important at small $Q^2$ ($\approx$ 1 GeV$^2$)
and at small $x$ as shown in Fig. 2.
Furthermore, the figure indicates that
our $Q^2$ variations are consistent with the CDHSW experimental data.
 } \ \hspace{0.8cm} \
\parbox{6.0cm}
{
\vspace{1.0cm}
\epsfile{file=fig2.eps,width=5.0cm}
\par\vspace{0.4cm}\hspace{-0.2cm}
Fig. 2  ~$Q^2$ evolution of $F_3$.
}

\vfill\eject
\noindent
{\bf 2.2 Singlet distribution}
\vspace{0.3cm}

The nonsinglet evolution is rather simple because
gluons do not participate in the evolution process.
However, the singlet part is more complicated
because the Altarelli-Parisi equations are coupled
integrodifferential equations in Eq. (1.2).
We do not discuss formalism of singlet solution, so
interested reader may read the papers in Ref. \cite{LAG}.
Even though the analysis is slightly complicated, the essential
procedure is the same as the one in
the previous subsection.
Our investigation for the singlet part is still in progress
\cite{KKK}, so the results in this subsection should be
considered as preliminary ones.
Calculating Laguerre-expansion coefficients for the splitting
functions in Eq. (1.2) and for initial distributions,
we obtain a solution similar to Eq. (2.5).

The initial singlet distribution $q_s(x)$ and gluon distribution $G(x)$
are chosen as the HMRS-B distributions at $Q^2$=4 GeV$^2$.
They are evolved to the ones at $Q^2$=200 GeV$^2$.
Three evolution results are shown by taking five, ten, and
twenty Laguerre polynomials in Figs. 3 and 4.
It is rather surprising to find excellent accuracy
even at very small $x$ as small as $10^{-4}$.
The accuracy is much better than the nonsinglet one
by taking only ten Laguerre polynomials.

\vspace{0.2cm}
\noindent
\parbox{7.0cm}
{
\vspace{0.4cm}
\epsfile{file=fig3.eps,width=7.0cm}
\par\vspace{0.3cm}\hspace{+1.0cm}
Fig. 3 ~$Q^2$ evolution of $xq_s(x)$.
 } \ \hspace{1.0cm} \
\parbox{7.0cm}
{
\vspace{0.4cm}
\epsfile{file=fig4.eps,width=7.0cm}
\par\vspace{0.3cm}\hspace{+1.0cm}
Fig. 4 ~$Q^2$ evolution of $xG(x)$.
}

$~~~$

$~~~$

Typical running time is a few seconds on SUN-IPX or on VAX-4000/500
in obtaining a $x$-distribution curve in Fig. 1, 3, or 4.
Considering the excellent accuracy and the short running time,
we conclude that the Laguerre method is very effective
for numerical solution of the Altarelli-Parisi equations.

\vfill\eject
\vspace{1.0cm}
\noindent
{\Large\bf 3. Brute-force method}
\vspace{0.4cm}

The Laguerre method is practically useful; however,
it does not produce excellent results
at small $x$ and at large $x$ in the nonsinglet case.
With development of high-energy accelerators, the small $x$
region becomes more and more interesting.
For those who are serious about the small $x$ and large $x$ regions,
we discuss an alternative method,
the brute-force method in Ref. \cite{BRUTE},
for obtaining accurate numerical solution.

The brute-force method is perhaps the simplest one
in solving the integrodifferential equation.
We divide the variables $t$ and $x$ into small steps, then
the derivative by $t$ is defined as
$$
{d \over {dt}} ~q(x,t)  ~ \Rightarrow ~
{{q(x_{i},t_{j+1}) - q(x_{i},t_{j})} \over {\Delta t}} ~~~,
\eqno{(3.1)}
$$
and the integral over $y$ is by
$$
\int_x^1 {{dy} \over y} ~P({x \over y}) ~q(y,t) ~\Rightarrow~
\sum_{k(=i)}^{N_x} {{\Delta x}\over{x_{k}}} ~
P({{x_{i}}\over {x_{k}}}) ~q(x_{k},t_j)  ~~~.
\eqno{(3.2)}
$$
If the number $N_x$ is increased, we expect to obtain
accurate numerical results.
However, numerical convergence is not good in the small
$x$ region even if we take large $N_x$.
So we decide to divide $x$ into equal steps in the logarithmic scale.
Then, our evolution results converge by taking
large $N_x \approx 200$.
Evolution results of the HMRS-B valence-quark
distribution are shown in Fig. 5
by taking $N_x$=50, 200, 500
with fixed number of points $N_t$=100 for the variable $t$.
The accuracy is excellent in the nonsinglet distribution,
so the brute-force method could be used in the small $x$ region
instead of the Laguerre method.
However, it takes significant amount of running CPU time
as shown in Fig. 6.

\vspace{0.2cm}
\noindent
\parbox{7.0cm}
{
\vspace{0.4cm}
\epsfile{file=fig5.eps,width=6.0cm}
\par\vspace{0.3cm}\hspace{0.5cm}
Fig. 5 ~$Q^2$ evolution of $x(u_v+d_v)$.
}
\ \hspace{1.4cm} \
\parbox{7.0cm}
{
\vspace{0.4cm}
\epsfile{file=fig6.eps,width=6.0cm}
\par\vspace{0.3cm}\hspace{0.5cm}
Fig. 6 ~Computing time.
}

\vfill\eject
\vspace{1.0cm}
\noindent
{\Large\bf 4. Conclusions}
\vspace{0.4cm}

We investigated numerical solution of the Altarelli-Parisi
equations with next-to-leading-order corrections
in the Laguerre-polynomial and the brute-force methods.
Excellent convergence is obtained in the singlet case by
using the Laguerre method with merely ten Laguerre polynomials.
The accuracy is slightly worse in the nonsinglet case especially
at small and large $x$.
However, accurate results are obtained in these regions
by using the brute-force method.

$~~~$

\vspace{0.2cm}

\begin{center}
{\Large\bf Acknowledgment} \\
\end{center}
\vspace{-0.2cm}

This research was partly supported by the Grant-in-Aid for
Scientific Research from the Japanese Ministry of Education,
Science, and Culture under the contract number 06640406.

$~~~$

\noindent
* Speaker

\noindent
$\star$ Our group activities are listed
at http://www.cc.saga-u.ac.jp/saga-u/riko/physics
/quantum1/structure.html
or at ftp://ftp.cc.saga-u.ac.jp/pub/paper/riko/quantum1.




\begin{thebibliography}{}
\bibitem{LAG}      W. Furmanski and R. Petronzio,
                       Nucl. Phys. {\bf B 195} (1982), 237;
                   G. P. Ramsey,
                     J. Comput. Phys. {\bf 60} (1985), 97;
                   S. Kumano and J. T. Londergan,
                       Comput. Phys. Commu. {\bf 69} (1992), 373.
\bibitem{KKK}      R. Kobayashi, M. Konuma, and S. Kumano,
                       preprint SAGA-HE-63-94
                       (to be published in Comput. Phys.
                        Commu.);
                   M. Konuma and S. Kumano, research in progress.
\bibitem{BRUTE}    M. Miyama and S. Kumano, research in progress.
\bibitem{HW}       R. T. Herrod and S. Wada,
                        Phys. Lett. {\bf 96B} (1980), 195.
\bibitem{HMRS}     P. N. Harriman, A. D. Martin, W. J. Stirling,
                   and R. G. Roberts,
                            Phys. Rev. {\bf D 42} (1990), 798.
\bibitem{NEUTRINO} P. Berge et al. (CERN-CDHSW collaboration),
                            Z. Phys. {\bf C 49} (1991), 187.
\bibitem{CCFR}     W. C. Leung et al. (Fermilab-CCFR collaboration),
                            Phys. Lett. {\bf B 317} (1993), 655.

\end{thebibliography}
\end{document}